\documentclass[showpacs,pra,twocolumn]{revtex4}
\usepackage{epsfig,amsmath}
\usepackage{subfigure}
\usepackage{graphicx}
\usepackage{dcolumn}
\usepackage{stmaryrd}
\usepackage{mathrsfs}
\usepackage{pifont}
\usepackage{amsthm}
\usepackage{amssymb}
\usepackage{bm}
\usepackage{latexsym}
\usepackage{color}
\usepackage{hyperref}
\usepackage{multirow}

\begin{document}

\title{Ten more times precision improved method for surface roughness estimation with weak measurement}
\author{Hui-Chao Qu$^{1}$, Ya Xiao$^{1}$ \footnote{Corresponding author: xiaoya@ouc.edu.cn}, Xin-Hong Han$^{1}$, Shan-Chuan Dong$^{1}$, and Yong-Jian Gu $^{1}$ \footnote{Corresponding author: yjgu@ouc.edu.cn}}
\address {$^{1}$College of Physics and Optoelectronic Engineering, Ocean University of China, Qingdao 266100, People's Republic of China}
\date{\today }
	
\begin{abstract}
High-precision surface roughness estimation plays an important role in many applications. However, the classical estimating methods are limited by shot noise and only can achieve the precision of 0.1 nm with white light interferometer. Here, we propose two weak measurement schemes to estimate surface roughness through spectrum analysis and intensity analysis. The estimating precision with spectrum analysis is about $10 ^{-5}$ nm by using a currently available spectrometer with the resolution of $\Delta \lambda= 0.04$ pm and the corresponding sensitivity is better than 0.1 THz/nm. And the precision and sensitivity of the light intensity analysis scheme achieve as high as 0.07 nm and 1/nm, respectively. By introducing a modulated phase, we show that the sensitivity and precision achieved in our schemes can be effectively retained in a wider dynamic range. We further provide the experimental design of the surface profiler based on our schemes. It simultaneously meets the requirements of high precision, high sensitivity, and wide measurement range, making it to be a promising practical tool.
	
\end{abstract}
\maketitle

\section{Introduction}
Workpieces with ultra-smooth surfaces are widely used in the fields of astronomy, aerospace, biomedicine and so on \cite{C2007,M2008,AY2013,ANT1995,ZSZ2009}. In order to accurately characterize the performance of the workpiece and provide technology guarantee for the manufacturing process, precise and reliable methods for surface roughness measurement are in high demand. And numerous techniques have been developed for this purpose, which can be divided into two basic types: contact \cite{YHZ2016,DDP2010} and non-contact \cite{TH1983,BWS2020}. Comparably, non-contact methods are advantageous due to the non-destructive nature and the ability to scan large areas in a short time \cite{FHP2015}. Most of non-contact methods are based on the light scattering \cite{BT1986} or interference \cite{C1979,HGZ2020} to obtain high precision. According to the latest research, the highest precision among these techniques is about 0.1 nm, which is achieved by white light interferometry \cite{JKK2014,HTS2020}. Though a flurry of studies have been carried out to improve the precision, the classical methods are theoretically bounded by the shot noise limit and diffraction limit. There is an urgent need to develop some advanced techniques to beat the classical bound.

Quantum weak measurement \cite{AAV1988, AV1990} can amplify tiny physical quantities \cite{SDJH2010,XKSVLG2013,LXTXG2012,HK2008,DSJH2009,SDWJH2010,LLHFLHZ2019,VMHFSDH2013} with the assistance of post-selection on the system under study, which becomes a powerful tool for precision measurement. The result of such post-selected assisted weak measurement can be expressed in terms of the weak value $A_{w}$ of the system observable $\widehat{A}$. When the system is pre-selected at $ \vert \psi_{i}\rangle $ and post-selected at $ \vert \psi_{f}\rangle $, this weak value is defined by
\begin{equation}
\begin{split}
&A_{w}=\dfrac{\langle \psi_{f}\vert \widehat{A}\vert \psi_{i}\rangle}{\langle \psi_{f} \vert \psi_{i}\rangle}.
\end{split}
\end{equation}
In general, it can be complex numbers. And it can become arbitrarily large when $ \vert \psi_{i} \rangle $ and $ \vert \psi_{f}\rangle $ are almost orthogonal. This is called weak value amplification (WVA). Numerous studies have shown that WVA can obtain higher precision than the conventional techniques \cite{BS2010,HBL2017,ZDW2015}, making it attract more and more attention.

However, considering a practical scenario, the measurement scheme should simultaneously satisfy the requirements of high precision, strong robustness, and wide dynamic range which can not be realized by the standard weak measurement (SWM). Great efforts have been made to improve the SWM performance: SWM with power-recycling \cite{WTH2016,KJG2021}, difference measurement \cite{HFZ2018,LQX2018}, or bias measurement \cite{ZCX2016,YZX2021} were introduced to further increase the measurement precision and robustness; inverse WVA  \cite{XLL2019} and adaptive WVA scheme \cite{LHZ2017} were developed to solve the problem of narrow high-sensitivity dynamic range that exists in the SWM. Though weak measurement can be carried out in many different systems, optical system is one of the best choices. The optical SWM is mostly presented for a single photon, which not only could be illustrated either quantum mechanically or classically, but also can yield an improved precision compared to classical methods. Recently, weak measurement has been widely accepted in precision metrology, but its application in surface roughness estimation was rarely reported. 

In this paper, following the strategies of Refs.~\cite{LQX2018, XLL2019, LHZ2017}, two weak measurement schemes are proposed for estimating the surface roughness of the workpiece with spectrum analysis and light intensity analysis, respectively.  The theoretical analysis illustrates that the measurement precision can exceed the optimal classical method by four orders of magnitude in the spectrum analysis and by one order of magnitude in the light intensity analysis. Both of them achieve high sensitivity. By adding a modulation strategy based on the feedback modulation of the spectrum shift and the intensity contrast ratio (ICR), we find that the achieved precision and sensitivity can be retained in an extended dynamic range. The experimental design of our schemes is also presented, which could be used for surface profile reconstruction.  

The paper is organized as follows. In Sec.\;\uppercase\expandafter{\romannumeral2}, we propose two theoretical schemes of surface roughness measurement with modulated weak measurement and give some numerical simulation results. Based on these schemes, we further present the corresponding experimental designs in Sec.\;\uppercase\expandafter{\romannumeral3}. Finally, we discuss the advantages of our schemes by comparing them with conventional methods and present a brief conclusion in Sec.\;\uppercase\expandafter{\romannumeral4}.
	
\begin{figure}[!h]%Fig.1
\centering\includegraphics[width=8cm]{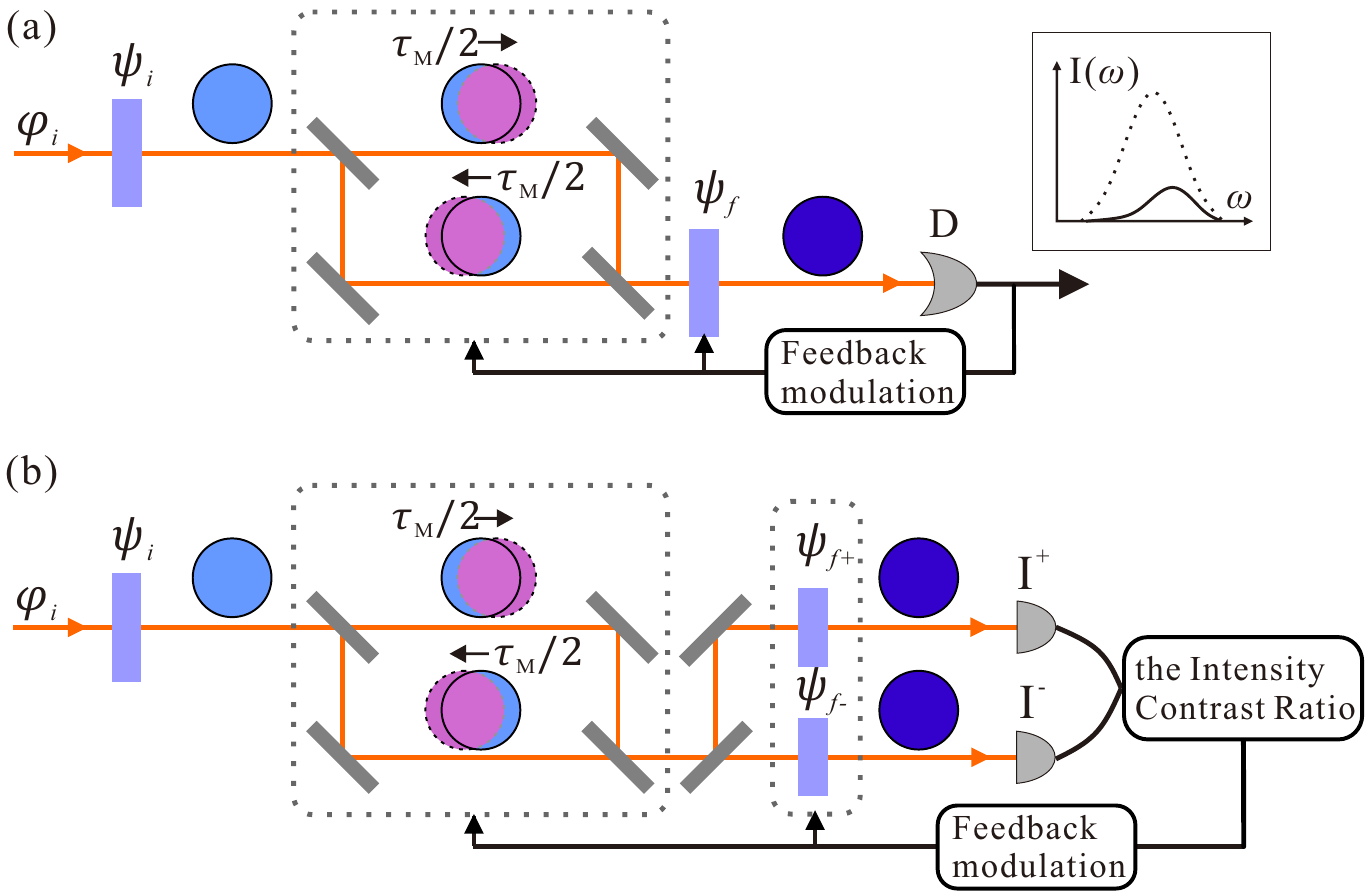}
\caption{(Color online) Schematic diagram of the surface roughness estimating schemes in (a) spectrum analysis and (b) light intensity analysis, which includes four steps: pre-selection, weak interaction, post-selection, and feedback modulation. The post-selection is adaptively adjusted via the feedback modulation from the spectrum shift and the ICR, respectively.}
\label{FIG1}
\end{figure}

\begin{figure*}[!t]%Fig.2
\centering\includegraphics[width=16cm]{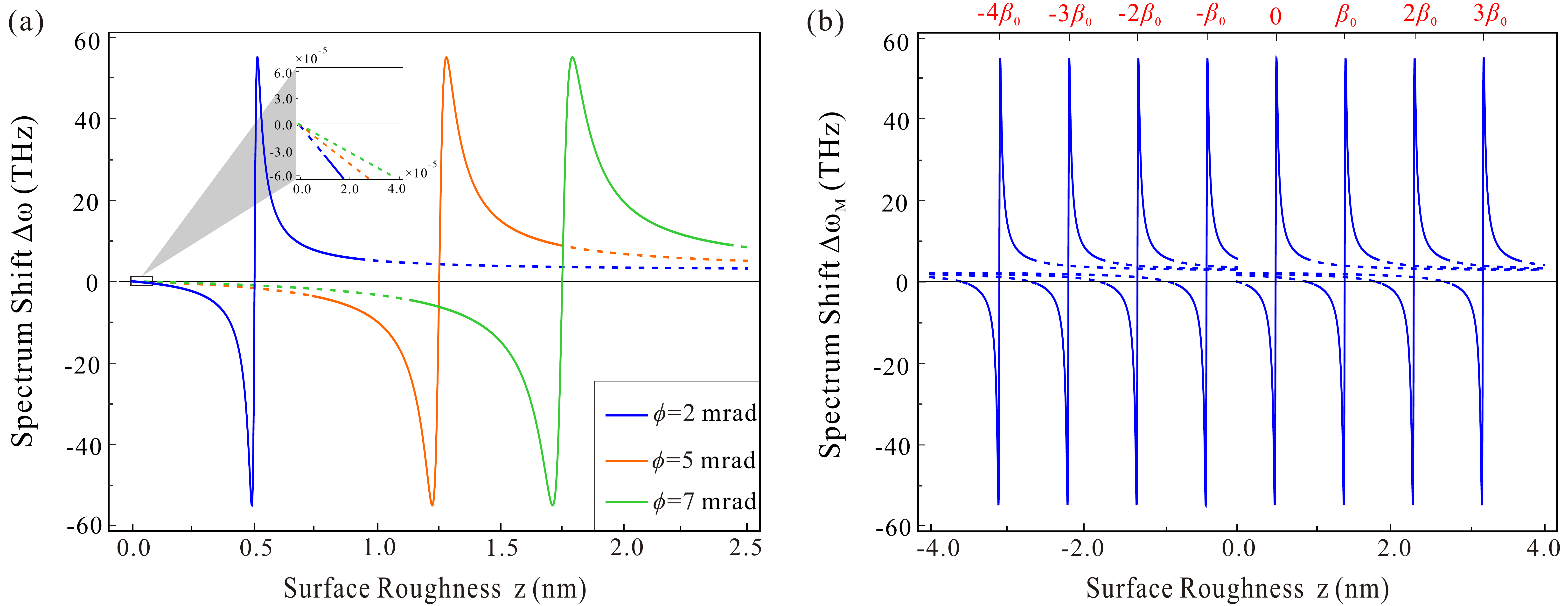}
\caption{(Color online) (a). The relation between the spectrum shift and the surface roughness without feedback modulations. The different colors correspond to different post-selected angles: $\phi=2$ mrad (blue), $ \phi =5$ mrad (orange), $ \phi=7 $ mrad (green). The solid parts of these curves denote the effective working ranges. In all three cases, their maximum spectral shifts are the same, making it possible to introduce feedback modulation to maintain high precision and sensitivity measurement. The inset shows that the minimum value that can be measured reaches about 10$ ^{-5} $ nm when $\phi= 2$ mrad. (b). The relation between the spectrum shift and the surface roughness with the feedback modulations when $\phi=2$ mrad. It clearly shows the high precision and sensitivity measurement range can be continuously extended.}
\label{FIG2}
\end{figure*}
\section{Theoretical Model}
%\label{Sec.2}
\textit{Modulated weak measurement with spectrum analysis}:
The theoretical schematic diagram of the surface roughness estimation based on weak measurement using spectrum analysis with the feedback modulation is shown in Fig.\;\ref{FIG1}(a). The estimating process comprises of four steps: pre-selection, weak interaction, post-selection, and feedback modulation. Here, we adopt the polarization and frequency of photon as the system and pointer. To perform the weak measurement, we need to pre-selected the system at a linear polarization state $\vert \psi_{i}\rangle=1/\sqrt{2}(\vert H\rangle+\vert V\rangle) $ \cite{ZDW2015,LHZ2017}, where $ \vert H\rangle $ and $ \vert V\rangle $ represent horizontal and vertical polarization as usual.

Then, the system is coupled weakly to a frequency-correlated Gaussian-shape pointer,  which is in the state $ \vert \varphi_{i}\rangle=\int d \omega f(\omega)\vert \omega\rangle $, where $ f(\omega)=(2 \pi\sigma^{2})^{-1/4}\exp[-(\omega-\omega_{0})^{2}/4\sigma^{2}] $. And the process is described by a interaction Hamiltonian  	
\begin{equation}\label{unitary}
\begin{split}
&\widehat{H}=g(t) \widehat{A}\otimes\widehat{\omega}, \\
\end{split}
\end{equation}
where $g(t) $ is the coupling strength, $\widehat{A}=\vert H\rangle\langle H\vert-\vert V\rangle\langle V\vert$ is the system observable. 

After the weak coupling process, the initial system-pointer joint state $ \vert \Psi_{i}\rangle=\vert \psi_{i}\rangle\otimes\vert \varphi_{i}\rangle $ evolves into
\begin{equation}\label{evolvedstate}
\begin{split}
\vert \Psi\rangle&=e^{-i\int\widehat{H}dt}\vert \psi_{i}\rangle \otimes \vert \varphi_{i}\rangle\\
&=e^{-i\int g(t) \widehat{A}\otimes\widehat{\omega}dt}\vert \psi_{i}\rangle \otimes \vert \varphi_{i}\rangle\\
&\simeq \dfrac{1}{\sqrt{2}}(\vert H\rangle e^{-i \omega \dfrac{z}{2c}} +\vert V\rangle e^{i\omega\dfrac{z}{2c}})\vert \varphi_{i}\rangle,\\
\end{split}
\end{equation}
under the first-order approximation with $ \vert z \sigma/c \vert\ll1 $, where $ \int g(t)dt=\tau/2=z/(2c) $. $ \tau $ is the time delay and $ c $ is the velocity of light. The surface roughness induces a phase shift $ \varphi $ between two orthogonal optical polarizations, which can be expressed as $\varphi=\omega z/c $. And the information of surface roughness $ z $ can be extracted by performing projective measurement on $ \vert \Psi\rangle $ \cite{KAN2012}. In order to realize a purely imaginary weak value to improve the measurement precision, the evolved state is post-selected on $ \vert \psi_{f}\rangle=1/\sqrt{2} (\vert H\rangle e^{-i\phi}-\vert V\rangle e^{i\phi}) $, where $ \phi $ is the post-selected angle. And the output pointer state collapses to 
\begin{equation}\label{pointerfinalstate}
\begin{split}
\vert \varphi_{d}\rangle&=\langle \psi_{f}\vert \Psi\rangle\\
&=\dfrac{-i}{\sqrt{P_{d}(z)}}\int d\omega f(\omega)\sin(\omega \dfrac{z}{2c}-\phi)\vert \omega\rangle, \\
\end{split}
\end{equation}
where $ P_{d}(z)=1/2(1-e^{- \sigma^{2}z^{2}/2c^2})\cos(\omega_{0} z/c-2\phi)$ denotes the successful post-selection probability. By measuring the output pointer, the unnormalized frequency probability distribution is obtained as follows:
\begin{equation}\label{distribution}
\begin{split}
&F(\omega)=\vert \langle \omega\vert \varphi_{d}\rangle \vert^{2}=\vert f(\omega)\vert^{2}\sin^{2}(\omega \dfrac{z}{2c}-\phi). \\
\end{split}
\end{equation}
And the relation between the mean spectrum shift and the surface roughness can be expressed as
\begin{equation}\label{spectrumshift}
\begin{split}
\Delta\omega&=\dfrac{\int \omega F(\omega)d\omega}{\int F(\omega)d\omega}-\omega_{0}\\
&= \dfrac{\sigma^{2}z}{2P_{d}(z)c} e^{- \sigma^{2}z^{2}/2c^{2}}\sin(\omega_{0} \dfrac{z}{c}-2\phi).\\
\end{split}
\end{equation}

Fig.\;\ref{FIG2}(a) shows the relation between the spectrum shift $ \Delta\omega $ and the surface roughness $ z $ under $ \phi= 2$, 5, 7 mrad, respectively. It can be seen that the spectrum shift changes its sign at the point $ z=2 \phi c/\omega_{0}$ and there is a sharp change of the spectrum shift around this point. Here, we choose the solid parts of these curves as the effective measuring ranges with the sensitivity higher than 0.1 THz/nm. It clearly shows that the smaller post-selected angle, the higher precision and sensitivity, and the narrower effective measuring range. Thus, a smaller post-selected angle should be chosen to ensure the more accurate measurement of the surface roughness. Considering the trade-off relation, we analyze the precision with $\phi= 2$ mrad, $\omega_{0}=2400 $ THz and $\sigma= 55$  THz. The precision is about 10$ ^{-5} $ nm by using a currently available spectrometer with the resolution of $\Delta \lambda= 0.04$ pm \cite{LLHFLHZ2019}, which is about four orders of magnitude better than the optimal classical method. However, the effective measuring range is restricted to 0.9 nm.

In order to extend the measuring range with high precision and sensitivity, we introduce a modulated phase $ \beta=\omega\tau_{M} $ in the feedback modulation of the spectrum shift, where $ \tau_{M} $ denotes the extra time delay \cite{HFZ2018}. And the total phase shift becomes $ \varphi_{t} =\varphi+\beta =\omega z/c + \omega \tau_{M}$. Hence the point at which the sign of the spectrum shift changes becomes $z=(2\phi+\beta)c/\omega_{0} $. The weak coupling process is rewritten as $ e^{-i(z/2c-\tau_{M})\widehat{\omega}\otimes\widehat{A}} $. Thus, the shift of the mean spectrum can be rewritten as
 \begin{small}
\begin{equation}\label{spectrumshiftM} 
\begin{split}
&\Delta\omega_{M} =\dfrac{\sigma^{2}(z/c -\tau_{M})e^{- \sigma^{2}(z/c-\tau_{M})^{2}/2}\sin(\omega_{0} z/c -2\phi-\beta)}{[1-e^{-\sigma^{2}(z/c-\tau_{M})^{2}/2}]\cos(\omega_{0}z/c-2\phi-\beta)}.\\
\end{split}
\end{equation}
\end{small}

The spectrum shift as a function of the surface roughness with the modulation procedure for $ \phi=2 $ mrad is depicted in Fig.\;\ref{FIG2}(b). Clearly, the effective surface roughness range can be doubled by adding the minimum unit modulated phase $ \beta_{0}=\omega_{0}z_{0}/c $, where $ z_{0} $ represents the minimum effective measurement range. That is to say, by sequentially increasing $ \beta_{0} $, e.g. $-4\beta_{0}$, $-3\beta_{0}$, $-2\beta_{0}$, $-\beta_{0}$, $0$, $\beta_{0}$, $2\beta_{0}$, $3\beta_{0} $, the spectrum shift yields periodic oscillation with the increase of the surface roughness and the surface roughness measurement range can be continuously extended with high precision and sensitivity.

\begin{figure*}[!t]%Fig.3
\centering\includegraphics[width=16cm]{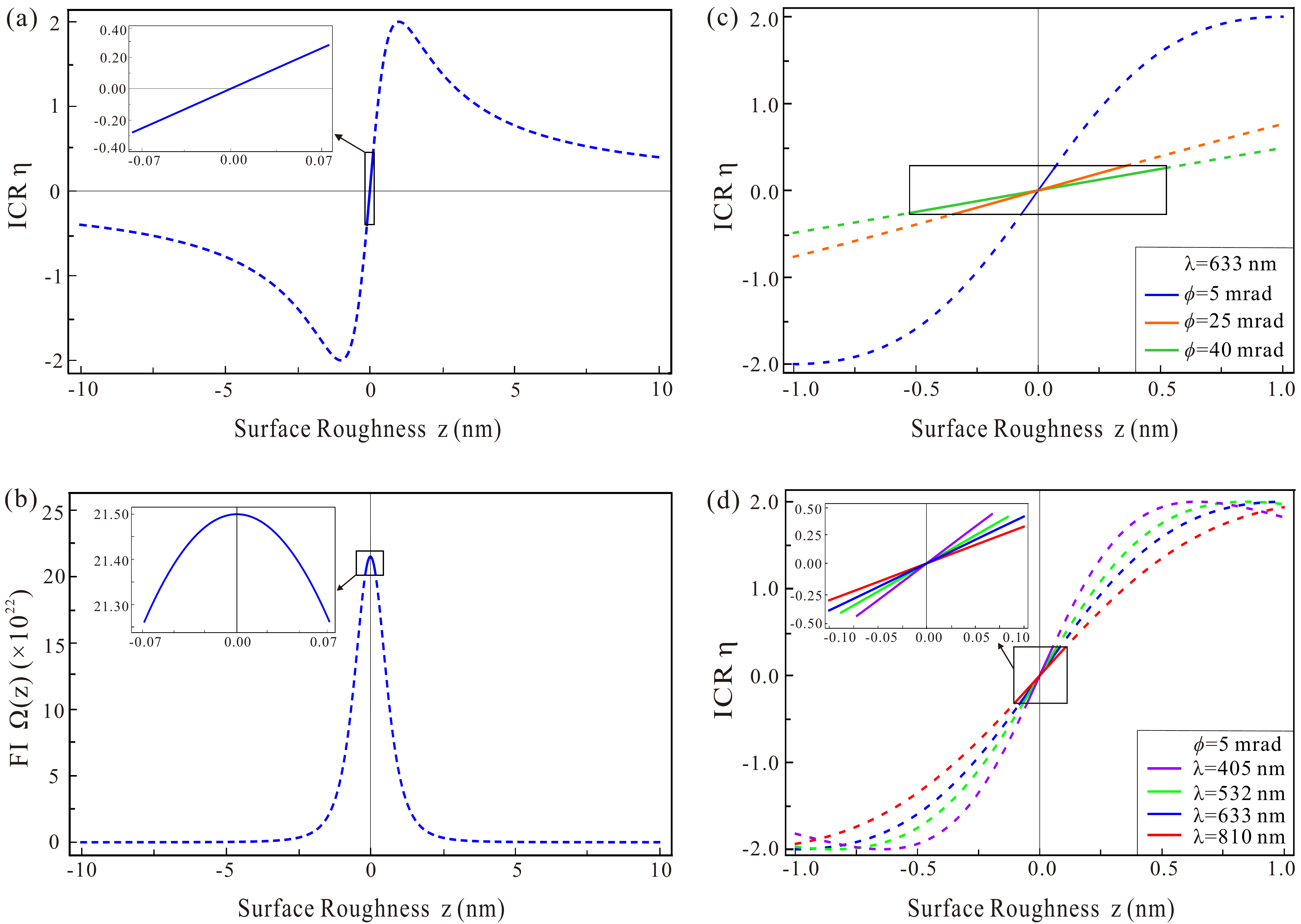}
\caption{(Color online) (a) ICR and (b) FI with respect to the surface roughness. The solid parts of these curves represent the approximate linear WVA range. (c) and (d) show the relation between ICR and the surface roughness $ z $ when the post-selected angle is chosen: $\phi= 5$ mrad (blue), $ \phi =25$ mrad (orange), $ \phi =40$ mrad (green) and the light wavelength is set at: $\lambda= 405$ nm (purple), $\lambda=532$ nm (green), $ \lambda=633 $ nm (blue), $ \lambda =810$ nm (red).}
\label{FIG3}
\end{figure*}

\textit{Modulated weak measurement with light intensity analysis}: The schematic diagram of the surface roughness estimation via weak measurement using light intensity analysis is depicted in Fig.\;\ref{FIG1}(b). Differently from the spectrum analysis method, the total post-selected light intensity is used to estimate the surface roughness in this scheme \cite{LQX2018}. We consider the same pre-selection state $ \vert \psi_{i}\rangle $ and initial pointer state $ \vert \varphi_{i}\rangle $ that are exploited in previous method. Similarly, the time delay $ \tau $ caused by the surface roughness is introduced by the weak interaction. The difference is that the evolved state is projected on two post-selection states $\vert \psi_{f\pm}\rangle $ with symmetric post-selected angles $ \pm\phi $, which are given as
$ \vert \psi_{f\pm}\rangle=(\vert H\rangle e^{\mp i\phi}-\vert V\rangle e^{\pm i\phi})/\sqrt{2} $. Meanwhile, the total post-selected light intensity $ I^{+} $ corresponding to post-selection state $ \vert \psi_{f+}\rangle $ can be derived explicitly
\begin{equation}\label{intensityfirst}
\begin{split}
I^{+}&=I_{0}\vert\langle\psi_{f+}\vert\Psi\rangle\vert^{2} \\
&=I_{0}(1-e^{-\sigma^{2}z^{2}/2c^2})\cos(\omega_{0} z/c -2\phi) \\
&\simeq I_{0}\sin^{2}(\omega_{0} z/2c-\phi) \\
&\simeq I_{0} \sin^{2}\phi[1-\textit{Im}(A_{\textit{w}})\omega_{0} z/c],\\
\end{split}
\end{equation}
where $ I_{0} $ denotes the light intensity without post-selection, and $\textit{Im}(A_{\textit{w}})=-\cot\phi $ represents the imaginary part of the weak value. The first approximation in Eq.\;(\ref{intensityfirst}) is established when $ \sigma $ is extremely small. The second approximation holds for the linear range $ \vert \omega_{0}z/c\vert/2\ll\vert\phi\vert $. Similarly, the total post-selected light intensity $ I^{-} $  can be expressed as $ I^{-}\simeq I_{0}(\sin^{2}\phi)[1+\textit{Im}(A_{\textit{w}})\omega_{0}z/c] $. To directly quantify the effect of  weak value amplification (WVA), we introduce the intensity contrast ratio (ICR) to normalize the intensity variation of the post-selected light caused by the surface roughness, which is defined as
\begin{equation}\label{ICR}
\begin{split}
&\eta=\dfrac{I^{+}-I^{-}}{(I^{+}+I^{-})/2}\simeq2\omega_{0}\dfrac{z}{c}\vert A_{\textit{w}}\vert, \\
\end{split}
\end{equation}
where $ \vert\eta\vert\leq 2 $. Obviously, the surface roughness can be directly amplified by a factor of $ \vert A_{\textit{w}}\vert\simeq1/\phi $ in the so-called linear WVA range when $ \vert\phi\vert\ll 1 $.

We plot the ICR curve with respect to the surface roughness in Fig.\;\ref{FIG3}(a). The ICR curve is always symmetric at $ z=0$. The linear WVA range (solid part) represents the high-sensitivity range, whose sensitivity $ {\partial \eta}/{\partial z} $ (the slope of the ICR curve) reaches 1/nm.
To clearly characterize the performance of the measurement scheme, we introduce the Fisher information (FI) to quantify the maximum achievable information, which can be expressed as
\begin{equation}\label{FI}
\begin{split}
&\Omega(z)=\partial^{2}_{z}\ln I(z)/I(z), \\
\end{split}
\end{equation}
where $I(z)$ is the total light intensity introduced by $\pm\phi$. 

As shown in Fig.\;\ref{FIG3}(b), the maximum FI is located at the peak of the curve, which is corresponding to the highest precision and sensitivity in the linear WVA range $\vert \omega_{0}z/c\vert/2\ll\vert\phi\vert $ with $ \phi= 5$ mrad. In addition, Fig.\;\ref{FIG3}(c) and (d) clearly show the smaller post-selected angle and the shorter wavelength, the narrower linear WVA range, the higher sensitivity and precision. Considering the trade-off relation, we analyze the precision with the $\phi= 5$ mrad and $ \lambda_{0}=633 $ nm. Here, we define measurement precision as the  measurable maximum surface roughness among the linear WVA range, which reaches 0.07 nm. It is about one order of magnitude higher than the optimal classical measurement method.

\begin{figure}[!ht]
\centering\includegraphics[width=8cm]{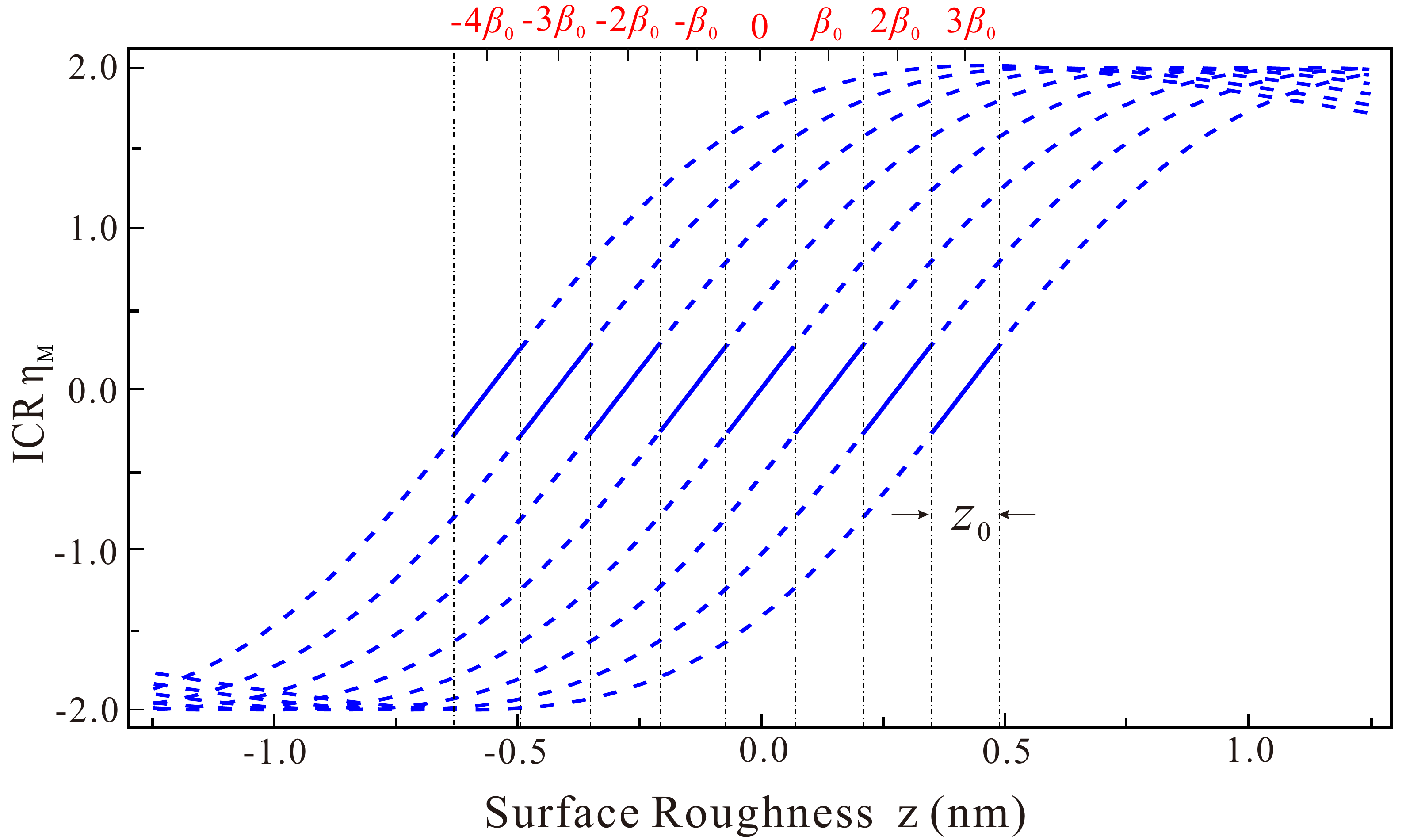}
\caption{(Color online) ICR as a function of the surface roughness by introducing different modulated phases. The solid parts of these curves represent the linear WVA range, which can be extended with retaining the high precision and sensitivity.}
\label{FIG4}
\end{figure}

However, there is an unavoidable problem that the dynamic range of the surface roughness estimation is limited by the fixed post-selected angle. Like the spectrum analysis method, the modulated procedure according to the feedback of the ICR is considered so as to significantly broadening the surface roughness estimating range without loss of the measurement precision and sensitivity \cite{HFZ2018,LQX2018}. Considering the same post-selection and the modulation procedure as above for the system, Eq.\;(\ref{ICR}) is modified to
\begin{equation}\label{ICRMWM}
\begin{split}
&\eta_{M}=\dfrac{I^{+}_{M}-I^{-}_{M}}{(I^{+}_{M}+I^{-}_{M})/2}\simeq2(\omega_{0}\dfrac{z}{c}-\beta)\vert A_{\textit{w}}\vert. \\
\end{split}
\end{equation}
The approximation is valid in the modulated linear WVA range $\vert\omega_{0}z/c-\beta\vert/2\ll\vert\phi_{M}\vert $. Fig.\;\ref{FIG4} shows that a wider surface roughness estimating range based on the modulated linear WVA range can be obtained by introducing $ \beta_{0}$ in turn.

\section{Experimental designs}
\begin{figure}[!h]%fig.5
\centering\includegraphics[width=8cm]{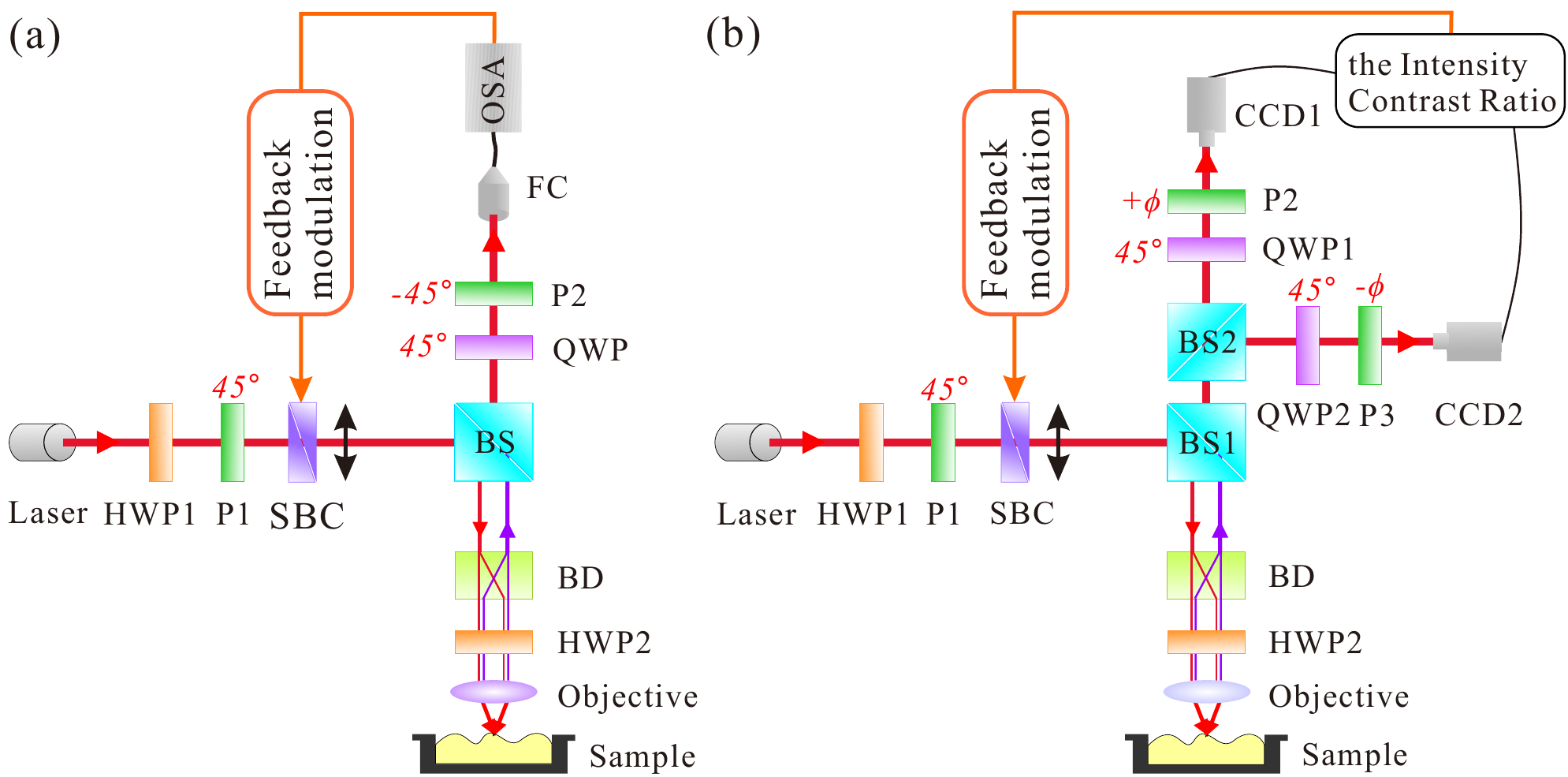}
\caption{(Color online) Schematic representation of the experimental setups for the surface roughness estimation based on weak measurement in (a) spectrum analysis and (b) light intensity analysis.}
\label{FIG5}
\end{figure}

We also present the experimental designs corresponding to the surface roughness estimation based on weak measurement in spectrum analysis and light intensity analysis, shown in Fig.\;\ref{FIG5}(a) and (b). The polarization of photon acts as the degree of freedom of system. In both schemes, a half-wave plate (HWP1) is used for adjusting the light intensity from the laser and the sets of polarizers (P1, P2) and quarter-wave plates (QWP1) are used for the pre- and post-selection. The time delay is introduced by calcite beam displacer (BD), HWP2, microscope objective and sample. The feedback modulation operation of the spectrum shift and the ICR can be realized by manually modulating the Soleil-Babinet compensator (SBC). In the first scheme, the fiber coupler (FC) collects the output beam whose spectrum is measured by using an optical spectrum analyzer (OSA). In the second scheme, the light intensity of the output beam is detected on charge-coupled devices (CCDs).

In the above two schemes, the spectrum shift and the ICR serve as the feedback information to timely adjust the pre- and post-selection with the increase of the surface roughness by sequentially adding a modulated phase. Since the ICR is linearly proportional to the surface roughness, the surface roughness estimation technique with light intensity analysis can also provide an advantage of reflecting the surface profile of the workpiece. We describe a set of the surface profile measurement procedure as follows:

(1) Perform the weak measurement in light intensity analysis with a set of initial post-selected angles, e.g. $ \phi=5 $ mrad, then the initial ICR $ \eta $ can be obtained.

(2) Adjust the modulated angle based on $ \eta $. If $\eta>\eta_{max} $, $ \beta_{0} $ will be added by adjusting the SBC; if $ \eta<\eta_{min} $, the modulated angle will be decreased by $ \beta_{0} $. Then perform the measurement again with the updated modulated angle. $ \eta_{max} $  
$(\eta_{min})$ is the maximum (minimum) ICR in the linear WVA range without the modulated phase. 

(3) Repeat above procedures until the $ \eta $ satisfies the linear WVA range $\eta_{min} \leq\eta\leq\eta_{max} $, and the point roughness of the surface can be extracted from the ultimate ICR. With point-by-point scanning the surface of the workpieces, we can directly reconstruct the surface profile by calculating the ICR and the introduced modulated angles.

\section{discussion and conclusion}
In summary, we present a new method to estimate the surface roughness of the workpieces based on weak measurement. We further discuss its performance in precision, sensitivity, and dynamic range from the view of practical application. Our results show that the measurement precision of the surface roughness estimation reaches $ 10^{-5} $ nm in spectrum analysis and reaches 0.07 nm in light intensity analysis, both better than the classical methods. The sensitivity of these analytical methods achieves 0.1 THz/nm and 1/nm, respectively, and it can be conveniently retained in an extended range by adding a feedback modulation. In theory, the noise introduced by the feedback modulation can be neglected \cite{HFZ2018, LQX2018}. The corresponding feasible experimental designs are also given. Based on our schemes, we also propose an improvement optical profiler design and give the corresponding operation procedures. Our precision improved schemes are valid for classical light, which can be easily implemented in practice. 

Compared with white light scanning interferometry, we can obtain the surface roughness by measuring the spectrum shift or the intensity contrast ratio, rather than analyzing the interference fringes. Therefore, the methods based on weak measurement can reduce the errors caused by optical effects such as diffraction, reflection and dispersion. What's more, our method can further beat the standard quantum limit if the quantum resources or techniques, such as quantum entanglement \cite{PDB2014}, nonlinearity \cite{FXS2011, CAS2018} and power recycle \cite{ WTH2016, KJG2021} are employed. Our work opens a new way for improving the sensitivity and accuracy in the field of surface roughness estimation.

\section{Acknowledgments}
This work was supported by the National Natural Science Foundation Regional Innovation and Development Joint Fund (Grant No. 932021070), the National Natural Science Foundation of China (Grants No. 912122020 and No. 61701464), the China Postdoctoral Science Foundation (Grant No. 861905020051), the Fundamental Research Funds for the Central Universities (Grants No. 841912027, and No. 842041012), the Applied Research Project of Postdoctoral Fellows in Qingdao (Grant No. 861905040045), and the Young Talents Project at Ocean University of China (Grant No. 861901013107).

\section*{Appendix:  Principle of white light scanning interferometry}

Various techniques are available for surface roughness measurement. Among these methods, white light scanning interferometry has been widely used in industry because of its non-destructive, large measurement range, high vertical resolution and high efficiency.

\begin{figure}[!h]%Fig.6
	\centering\includegraphics[width=9cm]{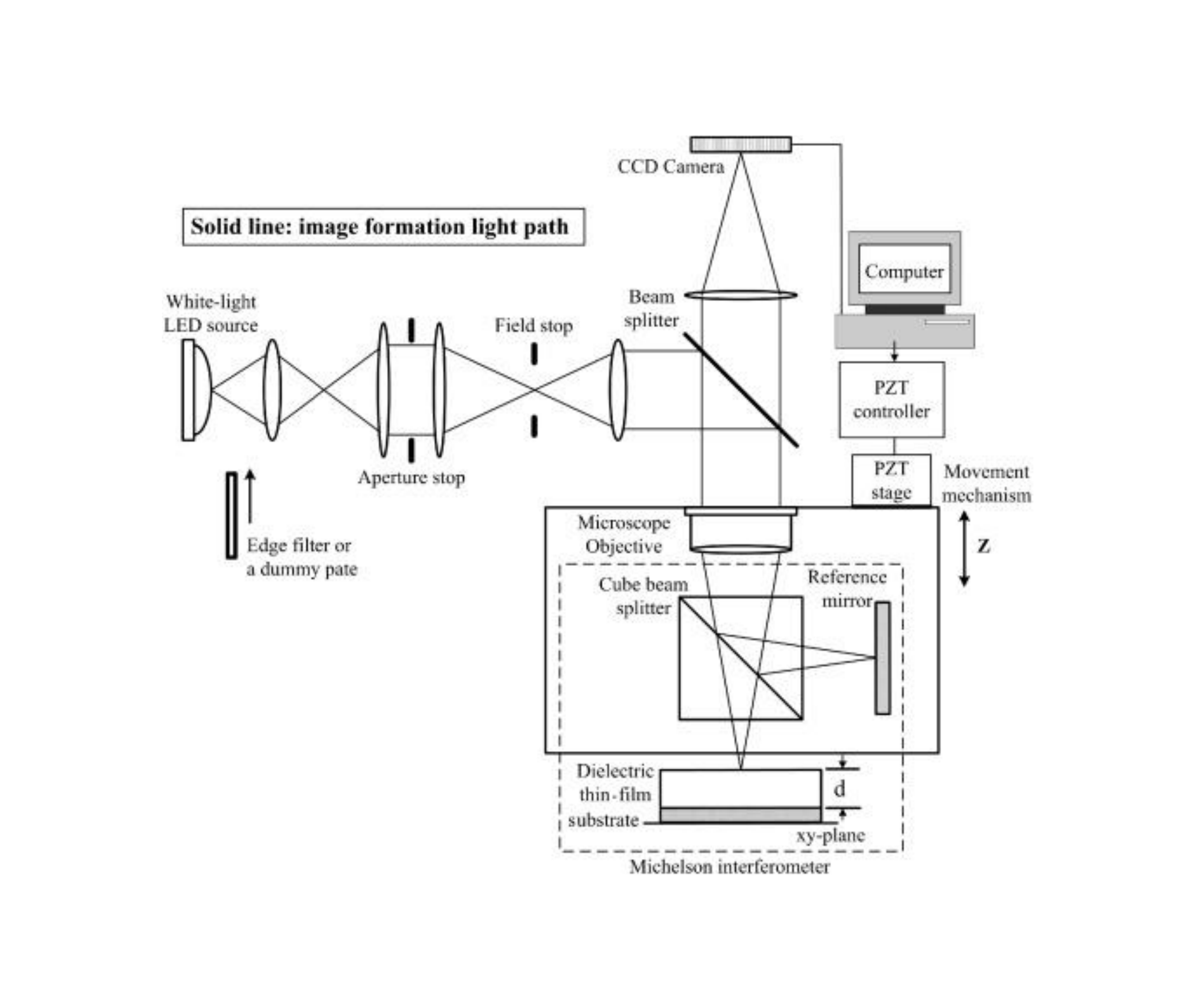}
	\caption{ Schematic of the white light scanning interferometry system \cite{LWL2012}.}
	\label{FIG6}
\end{figure}

A typical setup for white light scanning interferometry is shown in Fig.\;\ref{FIG6} \cite{LWL2012}. This device is mainly composed of light source, light reshape lens system, interferometry system, phase shifting system, CCD camera and analysis software. A broadband white light is filtered and expanded with a light reshape system, then collimated to a interferometry system. The system contains a beam splitter, which divides the incident beam into a measurement beam and a reference beam. The measurement beam passes to the sample surface, the reference beam is reflected and guided to a mirror. Light reflected from the sample and the mirror is recombined at the beam splitter and focused onto a CCD camera. The two reflected beams occur interfere if their optical path difference is within the coherence length of the illumination source. The asperity on the sample surface causes these path lengths to be unequal, which results in forming an interference pattern. As the sample surface is scanned with the PZT stage, the CCD camera records a series of interference patterns. With the analyse of the changes of the interference pattern of each scanned point, the three-dimensional surface morphology of the sample can be reconstructed.

However, since the surface roughness measurement with white light scanning interferometry is based on interference, the shape of interference fringes can easily be changed by diﬀraction, reflection and dispersion, such as the batwing effect \cite{AJ2000}, ghost steps \cite{AJ2001}, tilt-dependent dispersion \cite{P2003}, which will reduce the measurement accuracy and sensitivity. 

\end{document}